\documentclass[conference]{IEEEtran}
\IEEEoverridecommandlockouts
\usepackage{graphicx}
\usepackage{amsmath,amsthm}
\usepackage{amsfonts}
\usepackage{booktabs}
\usepackage{url}
\usepackage{tabularx}
\ifCLASSINFOpdf
\else
\fi
\begin{document}
\title{ Opportunistic DF-AF Selection Relaying with Optimal Relay Selection in Nakagami-$m$ Fading Environments}

\author{\IEEEauthorblockN{Tian Zhang{$^{*,\S}$}, Wei Chen$^{*}$, and Zhigang Cao$^{*}$}
\IEEEauthorblockA{$^{*}$State Key Laboratory on
Microwave and Digital Communications, Department of Electronic Engineering,\\
Tsinghua National Laboratory for Information Science and Technology
(TNList), Tsinghua University, Beijing, China.\\
$^{\S}$School of Information Science and  Engineering, Shandong University, Jinan, China.\\
Email: tianzhang.ee@gmail.com, \{wchen, czg-dee\}@tsinghua.edu.cn.}
\thanks{This work is partially supported by the National Nature Science Foundation of China under Grant No. 60832008 and No. 60902001.}}
\maketitle

\begin{abstract}
%\boldmath
%Recently, the opportunistic relaying technique
%%, which selects the best relay among potential relays to help a source
%%in sending its information to its destination,
% has gained much attention because of its simplicity and high spectral efficiency.
  An opportunistic DF-AF selection relaying scheme with maximal received signal-to-noise ratio (SNR) at the destination is investigated in this paper.
 The outage probability of the opportunistic DF-AF selection relaying scheme over Nakagami-$m$ fading channels is analyzed, and a closed-form solution is obtained. We perform asymptotic analysis of the outage probability in high SNR domain. The coding gain and the diversity order are obtained. For the purpose of comparison, the asymptotic analysis of opportunistic AF scheme in Nakagami-$m$ fading channels is also performed by using the Squeeze Theorem. In addition,
 we prove that compared with the opportunistic DF scheme and opportunistic AF scheme, the opportunistic DF-AF selection relaying scheme has better outage performance.
\end{abstract}


% \fi
\IEEEpeerreviewmaketitle

\section{Introduction}
\IEEEPARstart{C}{operative} diversity, which
lets the single
antenna equipped communication terminal enjoy the performance gain from spatial diversity, is an important modus operandi
of substantially improving coverage and performance in wireless
networks. The basic idea is that beside the direct transmission from the source to the destination, some adjacent nodes can be used to obtain the diversity by relaying the source signal to
the destination \cite{TC03: A. Sendonaris E. Erkip and B. Aazhang,IT04: J. N. Laneman D. N. C. Tse and G.W. Wornell}. Several cooperative diversity protocols including amplify-and-forward
(AF), decode-and-forward (DF), selection relaying and incremental relaying, have been discussed in \cite{IT04: J. N. Laneman D. N. C. Tse and G.W. Wornell}-\cite{TWC11:M. R. Bhatnagar and A. Hj phi rungnes}. DF-AF selection relaying protocol, where each relay can adaptively switch
between DF and AF according to its local signal-to-noise ratio (SNR), has been developed and investigated in \cite{Allerton03: B. Zhao and M. Valenti}-\cite{TC10: W. Su and X Liu}. \par
For the purpose of improving the system spectral efficiency,
the opportunistic relaying scheme for cooperative networks
has been introduced \cite{JSAC06:A. Bletsas A. Khisti D. P. Reed and A. Lippman,TWC08: E. Beres and R. Adve}. In such a scheme, a single relay is selected from a set of relay nodes. The opportunistic DF protocol and opportunistic AF protocol have been well studied in Rayleigh fading channels \cite{TWC07: A. Bletsas H. Shin and M.Z. Win}-\cite{TC10: S.S.Ikki and M.H.Ahmed}. In contrast, the opportunistic relaying over Nakagami-$m$ fading channels have not been extensively studied yet due to the mathematical difficulty. Furthermore,
when each relay utilizes DF-AF selection relaying protocol, how to choose a best one among them? \par
In this paper, we study the opportunistic DF-AF selection relaying with optimal relay selection whereby the destination obtains  maximal received SNR in cooperative networks. Moreover, we carry out asymptotic outage behavior analysis of the opportunistic DF-AF selection relaying scheme in Nakagami-$m$ fading channels. In addition, the asymptotic performance of opportunistic AF protocol in Nakagami-$m$ fading channels is analyzed for comparison.\par
The rest of the paper is structured as follows. Section II introduces the opportunistic DF-AF selection relaying scheme. Section III presents the asymptotic outage analysis of the opportunistic DF-AF selection relaying. Next, in Section IV,
comparisons with the opportunistic DF scheme and opportunistic AF scheme are performed. In Section V,
the numerical results are presented. Finally, the main results of the paper are summarized in Section VI.\par

% \fi
\IEEEpeerreviewmaketitle

\section{Description of opportunistic DF-AF selection relaying}
\begin{figure}[h]
\begin{center}
\includegraphics[width=3.5in]{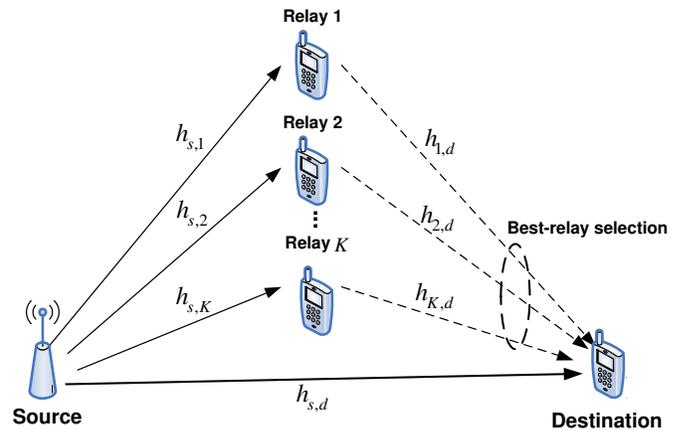}
\caption{Wireless relay channel}
\label{fig1}
\end{center}
\end{figure}
Consider a cooperative wireless network consisting of a source node ($S$), a destination node ($D$)
and $K$ potential relays $\mathbb{R}=(R_1,R_2,\ldots,R_K)$
as shown in Fig. \ref{fig1}.
Instantaneous SNRs of $S\rightarrow D$, $S\rightarrow R_i$ and $R_i\rightarrow D$ channels are denoted by $\gamma_{0}$, $\gamma_{1i}$ and $\gamma_{2i}$, where $i=1,2,\cdots,K$. The effects of the fading are captured by complex channel gains $h_{s,d}$, $h_{s,i}$ and $h_{i,d}$ respectively. The fading in each channel is assumed to be independent, slow and Nakagami-$m$ distributed with parameters $(m_0,\omega_0)$, $(m_{1i},\omega_{1i})$, and $(m_{2i},\omega_{2i})$ respectively. We have $\gamma_{0}=|h_{s,d}|^2 \mathsf{SNR}$, $\gamma_{1i}=|h_{s,i}|^2\mathsf{SNR}$, and $\gamma_{2i}=|h_{i,d}|^2 \mathsf{SNR}$, where $\mathsf{SNR}$ is the transmit SNR.\footnote{The destination and the relay apply equal transmit power $P$ unless otherwise specified, and $\mathsf{SNR}=P/N_{0}$ with $N_{0}$ being the AWGN variance at the relay and the destination.}\par
In opportunistic DF-AF selection relaying,
among the $K$-relay set, where each relay uses DF-AF selection relaying protocol, a \lq\lq best\rq\rq~relay will be selected for each transmission.
 Such a cooperative transmission is divided into two steps with equal durations. In the first step the source node broadcasts its signal to the destination node and the set of $K$-relay nodes as well. In the second step,
 the best relay $R_b$ is selected according to
\begin{equation}\label{bestselect}
b=\arg \max_{i=1,2,\cdots,K}\left\{\xi_i\gamma_{2i}+(1-\xi_i)\frac{\gamma_{1i}\gamma_{2i}}{\gamma_{1i}+\gamma_{2i}+1}\right\},
\end{equation}
where $\xi_i $ denotes the decoding state of $R_i$, i.e., $\xi_i =1$ when $R_i$ could fully decode the source message, otherwise $\xi_i =0$. And
 the best relay $R_b$ will use DF-AF selection relaying protocol to forward the received signal. Specifically, relay $R_b$ uses DF to re-transmit the received signal if $\xi_b =1$. Otherwise AF will be used. Since combing method does not affect the diversity order analysis, we assume that the destination uses Selection Combining (SC)
 to combine the signals received in the two stages for tractability unless otherwise specified.
%Fig. 2 presents the flow diagram of the transmission with the opportunistic DF-AF selection relaying scheme.
%\begin{figure}[h]
%\begin{center}
%\includegraphics[width=3.5in]{Flowchart.pdf}
%\caption{Flow diagram of two-stage transmission with opportunistic DF-AF selection relaying}
%\end{center}
%\end{figure}
\par
\emph{Remark:
%We consider relay selection together with relaying protocol selection in opportunistic DF-AF selection relaying.
 Let path $0$ represent the direct link ($S\rightarrow D$) and path $i$ represent the $i_{}^{th}$ cascaded link ($S\rightarrow R_i\rightarrow D$) where $i=1,\cdots,K$. For the $i_{}^{th}$ cascaded link we introduce a random variable $\gamma_{i}$ that will denote the equivalent instantaneous SNR at the destination \cite{TC10: S.S.Ikki and M.H.Ahmed}. That is, let $\gamma_{i}$ takes into account both the fading on $S \rightarrow R_i$ link and the fading on the $R_i \rightarrow D$ link. $\gamma_{i}=\gamma_{2i}$ when $R_i$ uses DF, otherwise if AF is used by $R_i$, $\gamma_{i}=\frac{\gamma_{1i}\gamma_{2i}}{\gamma_{1i}+\gamma_{2i}+1}$ \cite{IT04: J. N. Laneman D. N. C. Tse and G.W. Wornell}. Then we have
\begin{equation} \label{eqivelent}
\gamma_i=\xi_i \gamma_{2i}+(1-\xi_i )\frac{\gamma_{1i}\gamma_{2i}}{\gamma_{1i}+\gamma_{2i}+1}.
\end{equation}
Combining (\ref{bestselect}) and (\ref{eqivelent}), it can be shown that the proposed selection method of the best relay gives the maximal equivalent instantaneous SNR of $S\rightarrow R_b\rightarrow D$ path.
On the other hand, as Selection Combining (SC) is used at the destination. The received SNR at the destination can be given by
%\begin{equation}
$
\gamma=\max\{\gamma_0,\gamma_b\}.
$
%\end{equation}
Hence the proposed selection method of the best relay gives the maximal received SNR at the destination.
}
\section{Outage analysis of the opportunistic DF-AF selection relaying scheme}

\theoremstyle{definition} \newtheorem{lemma}{Lemma}

The outage probability can be given by the following lemma.
\begin{lemma}
The outage probability of the opportunistic DF-AF selection relaying scheme can be given by
\begin{eqnarray}\label{outage probability DF-AF}
\lefteqn{
P_{out}=\bigg[1-\frac{\Gamma\left(m_0,\alpha_0\gamma_{th}\right)}{\Gamma\left(m_0\right)}\bigg]
\prod_{i=1}^{K}\Bigg[\frac{\Gamma\left(m_{1i},\alpha_{1i}\Delta\right)}{\Gamma\left(m_{1i}\right)}
\Bigg(1
}
\nonumber\\
&-&\frac{\Gamma\left(m_{2i},\alpha_{2i}\gamma_{th}\right)}{\Gamma\left(m_{2i}\right)}\Bigg)+
\Bigg(1-\frac{\Gamma\left(m_{1i},\alpha_{1i}\Delta\right)}{\Gamma\left(m_{1i}\right)}\Bigg)
\Bigg],
\end{eqnarray}
where $R$ is the transmission rate, $\Delta=2^{2R}-1$ and $\gamma_{th}$ is the SNR threshold at the destination.
$\alpha_{0}=\frac{m_{0}}{\overline{\gamma}_{0}}$, $\alpha_{1i}=\frac{m_{1i}}{\overline{\gamma}_{1i}}$, and $\alpha_{2i}=\frac{m_{2i}}{\overline{\gamma}_{2i}}$, $\Gamma(\cdot)$ is the gamma function, and
  $\Gamma(\cdot,\cdot)$  is the incomplete gamma function.
\end{lemma}
\begin{IEEEproof}
Throughout this paper, we use  $X \sim \mathcal{G}(m)$ to denote that a random variable $X$ has the p.d.f. given by
$
f_{X}(y)=\frac{1}{\Gamma(m)}\left( \frac{m}{\overline{X}}\right)^{m}y^{m-1}e^ {-\frac{my}{\overline{X}}}.
$
First, it can be derived that c.d.f. of $Y \sim \mathcal{G}(m)$ can be given by
\begin{equation}\label{cdfgamma}
F_{Y}(y)=1-\frac{\Gamma\left(m,\frac{my}{\overline{Y}}\right)}{\Gamma\left(m\right)}.
\end{equation}
If the source-relay link is able to support $R$, i.e.,
$\frac{1}{2}\log_2\left(1+\gamma_{1i}\right)\ge R,$\footnote{The mutual information between $S$ and $R_i$ is given by $\frac{1}{2}\log_2\left(1+\gamma_{1i}\right)$ \cite{TC10: S.S.Ikki and M.H.Ahmed}.}
or equivalently, if
$\gamma_{1i}\ge \Delta=2^{2R}-1,$
the relay could fully decode the source message.
%\end{IEEEproof}
Consequently, the instantaneous equivalent end-to-end SNR per symbol at the destination is
%\begin{equation} \label{Total SNR at destination}
$
\gamma=\max\left(\gamma_0 , \max_{i=1,2,\cdots,K}\left\{\xi_i \gamma_{2i}+(1-\xi_i)\frac{\gamma_{1i}\gamma_{2i}}{\gamma_{1i}+\gamma_{2i}+1}\right\}\right),
%\end{equation}
$
where
%$\xi=1$ when the relay uses DF protocol, and $\xi=0$ when AF is used, or equivalently,
%\begin{equation}\label{probAF}
$\Pr\{\xi_i=0\}=\Pr\{\gamma_{1i}<\Delta\}=F_{\gamma_{1i}}(\Delta)$
%\end{equation}
and
%\begin{equation}\label{probDF}
$\Pr\{\xi_i=1\}=1-\Pr\{\xi_i=0\}.$
%\end{equation}
The outage probability can be given by
%\begin{eqnarray}\label{out}
$
P_{out}=\Pr\left\{\gamma < \gamma_{th}\right\}
=\Pr\left\{\gamma_0 < \gamma_{th}\right\}
\prod_{i=1}^{K} \Pr\left\{ \xi_i \gamma_{2i}+(1-\xi_i)\frac{\gamma_{1i}\gamma_{2i}}{\gamma_{1i}+\gamma_{2i}+1}<\gamma_{th}\right\}.$
%\end{eqnarray}
According to the Theorem of Total Probability and conditional probability, we have
\begin{eqnarray}  \label{proof of outage 2}
%\lefteqn{
P_{out}
&=& \Pr\{\gamma_0 <\gamma_{th}\}\prod_{i=1}^{K} \Big(\Pr\{\xi_i=1\}\Pr\{ \gamma_{2i}<\gamma_{th}\}
%}
\nonumber \\
&+&\Pr\{\xi_i=0\}\Pr\Big\{ \frac{\gamma_{1i}\gamma_{2i}}{\gamma_{1i}+\gamma_{2i}+1}<\gamma_{th}|\xi_i=0\Big\}\Big) \nonumber \\
&\stackrel{(a)}{=}& F_{\gamma_0}(\gamma_{th}) \prod_{i=1}^{K}\Big[(1-F_{\gamma_{1i}}(\Delta)) F_{\gamma_{2i}}(\gamma_{th})  \nonumber \\ &+&F_{\gamma_{1i}}(\Delta) \Big].
\end{eqnarray}
(a) holds since when $\xi_i=0$, i.e., $\gamma_{1i}< \Delta$, we have $\frac{\gamma_1\gamma_2}{\gamma_1+\gamma_2+1}<\gamma_{1i}<\Delta=\gamma_{th}$, i.e., $\Pr\Big\{ \frac{\gamma_{1i}\gamma_{2i}}{\gamma_{1i}+\gamma_{2i}+1}<\gamma_{th}|\xi_i=0\Big\}=1$.
Using (\ref{cdfgamma}) and (\ref{cdf AF}), we arrive at (\ref{outage probability DF-AF}).
\end{IEEEproof}
%The diversity order is given by the following theorem.
The following lemma presents the asympotic analysis (high $\mathsf{SNR}$) of the outage.
\begin{lemma}
%The diversity order is given by
%\begin{eqnarray}\label{div}
%d&=&-\lim_{\mathsf{SNR}\to \infty}\frac{\log P_{out}}{\log \mathsf{SNR}}
% \nonumber\\
% &=& m_0+\sum_{i=1}^{K}\min\{m_{2i},m_{1i}
%+\min\{m_{1i},m_{2i}\}\}.
%\end{eqnarray}
\begin{eqnarray}\label{asympotic DF}
  P_{out} \simeq \frac{\left(\frac{m_0\gamma_{th}}{\omega_0}\right)^{m_0}}{m_0\Gamma(m_0)}\mathsf{SNR}^{-m_0}
%  \nonumber\\
% &&
 \prod_{i=1}^{K}\Theta_i,
\end{eqnarray}
where $\simeq$ denotes asymptotic equality, and
\begin{eqnarray}
\Theta_i = \left\{
\begin{array}{ll}
\frac{1}{m_{2i}\Gamma(m_{2i})}\left(\frac{m_{2i}\gamma_{th}}{\omega_{2i}}\right)^{m_{2i}}\mathsf{SNR}^{-m_{2i}}, &  m_{1i}>m_{2i};\\
\frac{1}{m_{1i}\Gamma(m_{1i})}\left(\frac{m_{1i}\Delta}{\omega_{1i}}\right)^{m_{1i}}\mathsf{SNR}^{-m_{1i}}, &  m_{1i}<m_{2i};\\
\frac{2}{m_{1i}\Gamma(m_{1i})}\left(\frac{m_{1i}\Delta}{\omega_{1i}}\right)^{m_{1i}}\mathsf{SNR}^{-m_{1i}}        ,&  m_{1i}=m_{2i}.
\end{array} \right.
\end{eqnarray}

\end{lemma}
\begin{IEEEproof}
First, notice that the lower gamma function $\gamma(a,b) \simeq (1/a)b^a$ as $b \to 0$ \cite{GLOBECOM07: S. Savazzi and U. Spagnolini}.
%Let $ f(\mathsf{SNR})\sim \mathsf{SNR}^{d}$ denote $0<|\lim_{\mathsf{SNR}\to \infty}\frac{f(\mathsf{SNR})}{\mathsf{SNR}^{d}}|
% <\infty$.
 It can be shown that
 \begin{eqnarray} \label{1}
 \lefteqn{1-\frac{\Gamma\left(m_0,\alpha_0\gamma_{th}\right)}{\Gamma\left(m_0\right)}=\frac{\gamma\left(m_0,\alpha_0\gamma_{th}\right)}{\Gamma\left(m_0\right)}
}
 \nonumber\\
 &\simeq& \frac{1}{m_0\Gamma(m_0)}\left(\frac{m_0\gamma_{th}}{\omega_0}\right)^{m_0}\mathsf{SNR}^{-m_0}.
 \end{eqnarray}
 Similarly we obtain that
 \begin{eqnarray} \label{2}
 1-\frac{\Gamma\left(m_{1i},\alpha_{1i}\Delta\right)}{\Gamma\left(m_{1i}\right)}\simeq \frac{1}{m_{1i}\Gamma(m_{1i})}\left(\frac{m_{1i}\Delta}{\omega_{1i}}\right)^{m_{1i}}\mathsf{SNR}^{-m_{1i}}
 \end{eqnarray}
 and
  \begin{eqnarray}\label{3}
  \lefteqn{
  1-\frac{\Gamma\left(m_{2i},\alpha_{2i}\gamma_{th}\right)}{\Gamma\left(m_{2i}\right)}
  }\nonumber\\
& \simeq& \frac{1}{m_{2i}\Gamma(m_{2i})}\left(\frac{m_{2i}\gamma_{th}}{\omega_{2i}}\right)^{m_{2i}}\mathsf{SNR}^{-m_{2i}}.
 \end{eqnarray}

Using (\ref{1}), (\ref{2}), (\ref{3}), and (\ref{outage probability DF-AF}), the theorem can be obtained.
\end{IEEEproof}

\theoremstyle{definition} \newtheorem{corollary}{Corollary}
\begin{corollary}\label{div}
The coding gain in high $\mathsf{SNR}$, $g$,\footnote{When $P_{out} \simeq g\mathsf{SNR}^{-l}$, $g$ and $l$ are called coding gain and diversity order respectively.} can be expressed as
\begin{eqnarray} \label{df and af respectively SNR gain}
g= \frac{\left(\frac{m_0\gamma_{th}}{\omega_0}\right)^{m_0}}{m_0\Gamma(m_0)}\prod_{i=1}^{K}\eta_i.
\end{eqnarray}
$\eta_i=\frac{1}{m_{2i}\Gamma(m_{2i})}\left(\frac{m_{2i}\gamma_{th}}{\omega_{2i}}\right)^{m_{2i}}$ if $m_{1i}>m_{2i}$. Otherwise, $\eta_i=\frac{1}{m_{1i}\Gamma(m_{1i})}\left(\frac{m_{1i}\Delta}{\omega_{1i}}\right)^{m_{1i}}$.
\end{corollary}
\begin{corollary}\label{div}
The diversity order can be given by
\begin{eqnarray}
d&=&-\lim_{\mathsf{SNR}\to \infty}\frac{\log P_{out}}{\log \mathsf{SNR}}
 \nonumber\\
 &=& m_0+\sum_{i=1}^{K}\min\{m_{2i},m_{1i}\}.
\end{eqnarray}
\end{corollary}

%\subsection{Subsection Heading Here}
%Subsection text here.
%
%% needed in second column of first page if using \IEEEpubid
%%\IEEEpubidadjcol
%
%\subsubsection{Subsubsection Heading Here}
%Subsubsection text here.
\emph{Remark:
Corollary \ref{div} reveals that the diversity order depends not only on the number of the relays, $K$, but also the fading parameters of the channels. Meanwhile, the diversity order can be non-integer according to Corollary \ref{div}.
}
\section{Comparison with the opportunistic AF relaying scheme and opportunistic DF relaying scheme}
%The outage probability for the opportunistic DF scheme can be given by
%%\begin{figure*}
%\begin{eqnarray}\label{outage probability DF}
%\lefteqn{
%P_{df}=\bigg[1-\frac{\Gamma\left(m_0,\alpha_0\gamma_{th}\right)}{\Gamma\left(m_0\right)}\bigg]
%\prod_{i=1}^{K}\Bigg[\frac{\Gamma\left(m_{1i},\alpha_{1i}\Delta\right)}{\Gamma\left(m_{1i}\right)}
%\Bigg(1
%}
%\nonumber\\
%&-&\frac{\Gamma\left(m_{2i},\alpha_{2i}\gamma_{th}\right)}{\Gamma\left(m_{2i}\right)}\Bigg)+
%\Bigg(1-\frac{\Gamma\left(m_{1i},\alpha_{1i}\Delta\right)}{\Gamma\left(m_{1i}\right)}\Bigg)
%\Bigg].
%%          &\times& \Pr\left\{\gamma_0 < \gamma_{th}\right\}\prod_{i=1}^{K} \Pr\{ \xi_i \gamma_{2i}+(1-\xi_i)\frac{\gamma_{1i}\gamma_{2i}}{\gamma_{1i}+\gamma_{2i}+1}<\gamma_{th}\}\nonumber \\
%\end{eqnarray}
%%\end{figure*}
First, c.d.f. of $\frac{\gamma_{1i}\gamma_{2i}}{\gamma_{1i}+\gamma_{2i}+1}$ is given by \cite{EGWCN06:. A. Tsiftsis G. K. Karagiannidis P. T. Mathiopoulos and S. A. Kotsopoulos}
\begin{eqnarray} \label{cdf AF}
\lefteqn{
F_{\frac{\gamma_{1i}\gamma_{2i}}{\gamma_{1i}+\gamma_{2i}+1}}(y)=
1-\frac{2\alpha_{2i}^{m_{2i}}(m_{1i}-1)!e^{-(\alpha_{1i}+\alpha_{2i})y}}{\Gamma(m_{1i})\Gamma(m_{2i})}
}
\nonumber \\
&\times&\sum_{n=0}^{m_{1i}-1}\sum_{j=0}^{n}\sum_{k=0}^{m_{2i}-1}\bigg[\frac{1}{n!}
{n\choose j}{m_{2i}-1 \choose k}
  \alpha_{1i}^{\frac{2n-j+k+1}{2}}
 \nonumber \\
 &\times&
 \alpha_{2i}^{\frac{j-k-1}{2}}(1+y)^{\frac{j+k+1}{2}}y^{\frac{2n+2m_{2i}-j-k-1}{2}}\nonumber \\
 &\times&K_{j-k-1}\left(2\sqrt{\alpha_{1i}\alpha_{2i}y(y+1)}\right)\bigg],
\end{eqnarray}
where $K_{v}(\cdot)$ denotes the $v^{th}$ order modified Bessel function of the second kind.
Consequently,
the outage probability for the opportunistic AF scheme is given by
%\begin{figure*}
\begin{eqnarray}\label{outage probability AF}
P_{af}&=&\Pr\left\{\max\left(\gamma_0 , \max_{i=1,2,\cdots,K}\left\{\frac{\gamma_{1i}\gamma_{2i}}{\gamma_{1i}+\gamma_{2i}+1}\right\}\right)<\gamma_{th}\right\}\nonumber\\
&=&\bigg[1-\frac{\Gamma\left(m_0,\alpha_0\gamma_{th}\right)}{\Gamma\left(m_0\right)}\bigg]
F_{\frac{\gamma_{1i}\gamma_{2i}}{\gamma_{1i}+\gamma_{2i}+1}}(y),
%          &\times& \Pr\left\{\gamma_0 < \gamma_{th}\right\}\prod_{i=1}^{K} \Pr\{ \xi_i \gamma_{2i}+(1-\xi_i)\frac{\gamma_{1i}\gamma_{2i}}{\gamma_{1i}+\gamma_{2i}+1}<\gamma_{th}\}\nonumber \\
\end{eqnarray}
%\end{figure*}
where $F_{\frac{\gamma_{1i}\gamma_{2i}}{\gamma_{1i}+\gamma_{2i}+1}}(y)$ is given by (\ref{cdf AF}).
Denote $\gamma_{i}=\min\{\gamma_{1i},\gamma_{2i}\}$, when $\mathsf{SNR} \to \infty$, observe that $\frac{1}{2}\gamma_{i}\leq\frac{\gamma_{1i}\gamma_{2i}}{\gamma_{1i}+\gamma_{2i}+1}<\gamma_{i}$ \cite{TWC04: P. A. Anghel and M. Kaveh}.
 Therefore we have
%\begin{eqnarray} \label{low&up}
%\lefteqn{
$
\Pr\left\{\gamma_{i} <\gamma_{th}\right\}
%} \nonumber \\
<\Pr\left\{\frac{\gamma_{1i}\gamma_{2i}}{\gamma_{1i}+\gamma_{2i}+1} <\gamma_{th}\right\}
% \nonumber \\
\leq  \Pr\left\{\frac{1}{2}\gamma_{i}<\gamma_{th}\right\}.$
%\end{eqnarray}
Using (\ref{cdfgamma}), it can be shown that
\begin{eqnarray} \label{lower asym}
\lefteqn{
\Pr\left\{\gamma_{i}<\gamma_{th}\right\}
}\nonumber \\
&=&1-\frac{\Gamma\left(m_{1i},\alpha_{1i}\gamma_{th}\right)\Gamma\left(m_{2i},\alpha_{2i}\gamma_{th}\right)}{\Gamma\left(m_{1i}\right)\Gamma\left(m_{2i}\right)}
\nonumber \\
&=&\Bigg(1-\frac{\Gamma\left(m_{1i},\alpha_{1i}\gamma_{th}\right)}{\Gamma\left(m_{1i}\right)}\Bigg)
+\Bigg(1
-\frac{\Gamma\left(m_{2i},\alpha_{2i}\gamma_{th}\right)}{\Gamma\left(m_{2i}\right)}\Bigg)\nonumber \\
&-&\Bigg(1-\frac{\Gamma\left(m_{1i},\alpha_{1i}\gamma_{th}\right)}{\Gamma\left(m_{1i}\right)}\Bigg)\Bigg(1
-\frac{\Gamma\left(m_{2i},\alpha_{2i}\gamma_{th}\right)}{\Gamma\left(m_{2i}\right)}\Bigg).
\end{eqnarray}
Combining (\ref{2}), (\ref{3}), and (\ref{lower asym}), it can be shown that
\begin{eqnarray} \label{lowdiv}
\Pr\left\{\gamma_{i}<\gamma_{th}\right\}\simeq \frac{1}{m_i\Gamma(m_i)}\left(\frac{m_i\gamma_{th}}{\omega_i}\right)^{m_i}\mathsf{SNR}^{-m_{i}}:=L_i,
\end{eqnarray}
where $m_i=\min\{m_{1i},m_{2i}\}$.
Similarly, we get
\begin{eqnarray} \label{updiv}
\Pr\left\{\frac{1}{2}\gamma_{i}<\gamma_{th}\right\}&\simeq& \frac{1}{m_i\Gamma(m_i)}\left(\frac{2m_i\gamma_{th}}{\omega_i}\right)^{m_i}\mathsf{SNR}^{-m_{i}} \nonumber \\
&:=&U_i.
\end{eqnarray}
%Combining (\ref{low&up}), (\ref{lowdiv}), and (\ref{updiv}),
Consequently, we have
$L_i \preceq \Pr\left\{\frac{\gamma_{1i}\gamma_{2i}}{\gamma_{1i}+\gamma_{2i}+1} <\gamma_{th}\right\}\preceq U_i,$
 i.e.,
\begin{eqnarray} \label{asym}
L_i \preceq F_{\frac{\gamma_{1i}\gamma_{2i}}{\gamma_{1i}+\gamma_{2i}+1}}(y) \preceq U_i,
 %\nonumber \\
% &\sim& \mathsf{SNR}^{-\min\{m_{1i},m_{2i}\}}.
\end{eqnarray}
where $f_1(\mathsf{SNR}) \preceq f_2(\mathsf{SNR})$ means $0<\lim\limits_{\mathsf{SNR} \to \infty}\frac{f_1(\mathsf{SNR})}{f_2(\mathsf{SNR})} \le 1$.
Using (\ref{1}), (\ref{2}), (\ref{3}), (\ref{asym}), we have
%\begin{eqnarray}\label{asympotic aF}
$
  P_{af} \simeq \frac{\left(\frac{m_0\gamma_{th}}{\omega_0}\right)^{m_0}}{m_0\Gamma(m_0)}\mathsf{SNR}^{-m_0}
%  \nonumber\\
% &&
 \prod_{i=1}^{K}\Lambda_i,
 $
%\end{eqnarray}
$L_i\preceq \Lambda_i \preceq U_i$.
Then, we obtain the coding gain
%\begin{eqnarray} \label{df and af respectively SNR gain}
$
g_{af}= \frac{\left(\frac{m_0\gamma_{th}}{\omega_0}\right)^{m_0}}{m_0\Gamma(m_0)}\prod_{i=1}^{K}\eta_i^{af}
$
with
%\end{eqnarray}
$\frac{1}{m_i\Gamma(m_i)}\left(\frac{m_i\gamma_{th}}{\omega_i}\right)^{m_i} \le \eta_i^{af} \le \frac{1}{m_i\Gamma(m_i)}\left(\frac{2m_i\gamma_{th}}{\omega_i}\right)^{m_i}$.
%(Recall that $m_i=\min\{m_{1i},m_{2i}\}$ )
and
the diversity order
%\begin{eqnarray} \label{df and af respectively div}
$
d_{af}= m_0+\sum_{i=1}^{K}\min\{m_{2i},m_{1i}\}.
$
%\end{eqnarray}
\par
\emph{Remark 1: It is difficult to perform the asymptotic analysis ($\mathsf{SNR} \to \infty$) directly based on (\ref{outage probability AF}). The difficulty is that the c.d.f. of the the equivalent relay path SNR $\frac{\gamma_{1i}\gamma_{2i}}{\gamma_{1i}+\gamma_{2i}+1}$ is very complicated.
In this paper, we solve the difficulty by bounding the the equivalent relay path SNR with simple lower and upper bounds in high SNR. Specifically,  $\frac{1}{2}\gamma_{i}\leq\frac{\gamma_{1i}\gamma_{2i}}{\gamma_{1i}+\gamma_{2i}+1}<\gamma_{i}$. Then we have the simple lower and upper bounds of the c.d.f. of the the equivalent relay path SNR. Moreover, the lower bound and the upper bound have the same $\mathsf{SNR}$ order. Consequently, we obtain the asymptotic behavior of the outage probability.
}
\par
\emph{Remark 2: The opportunistic DF-AF and opportunistic AF have the same diversity order, i.e., $d=d_{af}$. However,
when $\Delta=\gamma_{th}$, we have $g \le g_{af}$. That is to say, the opportunistic DF-AF has lower outage than opportunistic AF in high SNR.
}
%When $m_{1i}\ge m_{2i}$, $\min\{m_{2i},m_{1i}
%+\min\{m_{1i},m_{2i}\}\}=m_{2i}$ and $\min\{m_{2i},m_{1i}\}=m_{2i}$. We have $\min\{m_{2i},m_{1i}
%+\min\{m_{1i},m_{2i}\}\}=\min\{m_{2i},m_{1i}\}$; Otherwise when $m_{1i} < m_{2i}$, $\min\{m_{2i},m_{1i}\}=m_{1i}$ and $\min\{m_{2i},m_{1i}
%+\min\{m_{1i},m_{2i}\}\}=\min\{m_{2i},2m_{1i}\}=$\[\left\{ \begin{array}{ll}
%2{m_{1i}};&if ~2{m_{1i}} \le {m_{2i}}\\
%{m_{2i}};&else
%\end{array} \right.\]$>m_{1i}$. Then, we get
%$\min\{m_{2i},m_{1i}+\min\{m_{1i},m_{2i}\}\}>\min\{m_{2i},m_{1i}\}$.
%\par
%Consequently, $d \ge d_{af}=d_{df}$. Equality occurs only when for all $i=1,\cdots,K$, $m_{1i}\ge m_{2i}$.
%\par
%In conclusion, we have the following lemma
%\begin{lemma}\label{lemma div improvment}
%$d > d_{af}=d_{df}$ when $\exists i \in [1,K]$ satisfying $m_{1i}< m_{2i}$; Otherwise $d = d_{af}=d_{df}$.
%\end{lemma}

When SC is applied at the destination, the outage probability for opportunistic DF scheme is the same as opportunistic DF-AF.\footnote{The outage probability for opportunistic DF is also given by (\ref{outage probability DF-AF}).} The opportunistic DF scheme has the same outage behavior as the opportunistic DF-AF scheme in this case.\footnote{The diversity order and coding gain are the same consequently.} However, if Maximal Ratio Combining (MRC) is used, the outage probability for opportunistic DF-AF is given by
\begin{eqnarray}
\lefteqn{
P_{out}^{'}=\Pr\Bigg\{\gamma_0+\max_{i=1,2,\cdots,K}\left\{\gamma_i\right\}< \gamma_{th}\Bigg\}
}
 \nonumber\\
&=& \Pr\Bigg\{\max_{i=1,2,\cdots,K}\bigg\{\gamma_0+\gamma_i\bigg\}<\gamma_{th}\Bigg\}
\nonumber\\
&=&\prod_{i=1}^{K} \Big(\Pr\{\xi_i=1\}\Pr\{ \gamma_0+\gamma_{2i}<\gamma_{th}\}
+\Pr\{\xi_i=0\}
\nonumber \\
&&\Pr\Big\{ \gamma_0+\frac{\gamma_{1i}\gamma_{2i}}{\gamma_{1i}+\gamma_{2i}+1}<\gamma_{th}|\xi_i=0\Big\}\Big).
\end{eqnarray}
In contrast, the outage probability for opportunistic DF can be derived as
\begin{eqnarray}
\lefteqn{
P_{df}^{'}=\Pr\Bigg\{\gamma_0
+ \max_{i=1,2,\cdots,K}\left\{\xi_i\gamma_{2i}\right\}<\gamma_{th}\Bigg\}
}
\nonumber\\
&=&\prod_{i=1}^{K} \Big(\Pr\{\xi_i=1\}\Pr\{ \gamma_0+\gamma_{2i}<\gamma_{th}\}
+\Pr\{\xi_i=0\}
\nonumber \\
&&\Pr\Big\{ \gamma_0<\gamma_{th}|\xi_i=0\Big\}\Big)
\end{eqnarray}
Since $\Pr\Big\{ \gamma_0<\gamma_{th}|\xi_i=0\Big\}>\Pr\Big\{ \gamma_0+\frac{\gamma_{1i}\gamma_{2i}}{\gamma_{1i}+\gamma_{2i}+1}<\gamma_{th}|\xi_i=0\Big\}$, we have $P_{out}^{'}<P_{df}^{'}$.
On the other hand, the outage for opportunistic AF with MRC is
\begin{eqnarray}
P_{af}^{'}=\Pr\Bigg\{\gamma_0
+ \max_{i=1,2,\cdots,K}\left\{\frac{\gamma_{1i}\gamma_{2i}}{\gamma_{1i}+\gamma_{2i}+1}\right\}<\gamma_{th}\Bigg\}
\end{eqnarray}
Since $\frac{\gamma_{1i}\gamma_{2i}}{\gamma_{1i}+\gamma_{2i}+1}<\gamma_{2i}$, then $P_{out}^{'}<P_{af}^{'}$.
%%%%%%%%%%%%%%%
\section{Numerical results}
In this section, computer simulations are performed to verify
the accuracy of our derived analytical results.
In the simulations, we have set $R=1$ bit/sec/HZ, $\omega_0=\omega_{1i}=\omega_{2i}=1$. \par
%When $S-R_b$ in outage, $R_b$ use AF; Otherwise, $R_b$ decodes and forwards the received information. Compared with AF, there is extra decoding power consumption in DF.
\begin{figure}[h]
\begin{center}
\includegraphics[width=3.4in]{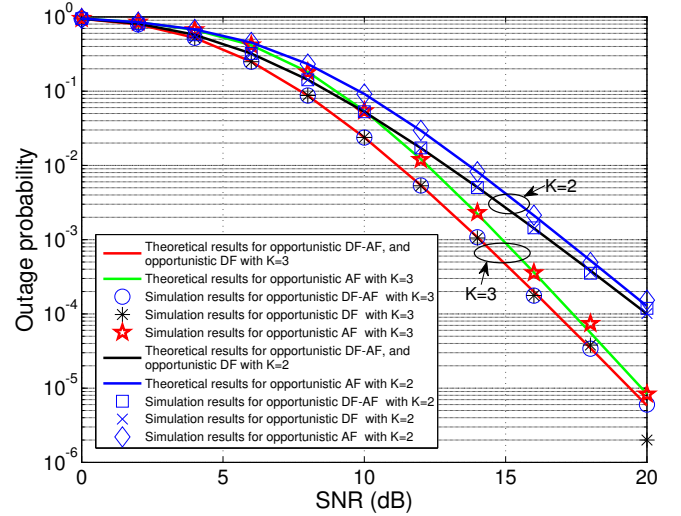}
\caption{Outage probability in Rayleigh channels with $K$ relays}
\label{fig2}
\end{center}
\end{figure}
\begin{figure}[h]
\begin{center}
\includegraphics[width=3.4in]{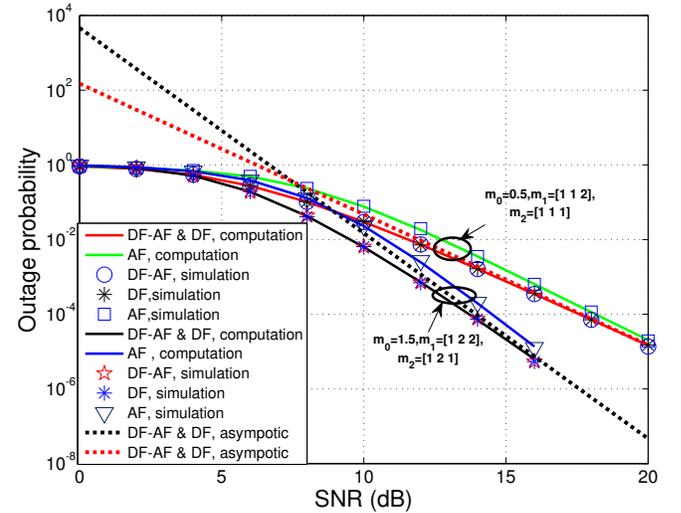}
\caption{Outage probability in Nakagami-$m$ channels with 3 relays}
\label{fig3}
\end{center}
\end{figure}
Fig. \ref{fig2} compares the outage probability obtained via simulations
and theoretical evaluation with different number of potential cooperating relays ($K$) in Rayleigh fading environment ($m_0 =
m_{1i} = m_{2i} = 1$). Fig. \ref{fig3} compares the outage probability with general Nakagami-$m$ fading parameters in $K=3$ relays scenario, $m_0=0.5,m_{1}=[1 ~1 ~2],m_{2}=[1 ~1 ~1]$.
As a benchmark, we
also show the outage probability of the best-relay selection adaptive DF scheme \cite{TC10: S.S.Ikki and M.H.Ahmed}
as well as the outage probability of the opportunistic AF schem. Observe that simulation
curves match in high accuracy with analytical ones. When SC is utilized, opportunistic DF-AF scheme has the same outage probability
as the best-relay selection adaptive DF scheme, which has better outage performance than the opportunistic AF scheme. The asymptotic outage coincides with the exact outage in high SNR region. We can notice that both the number of potential cooperating relays
($K$) and the fading parameters have a strong impact of the
performance enhancement. We will analyze the impact of the relay number and the fading parameter respectively in the following.

\par
\begin{figure}[h]
\begin{center}
\includegraphics[width=3.4in]{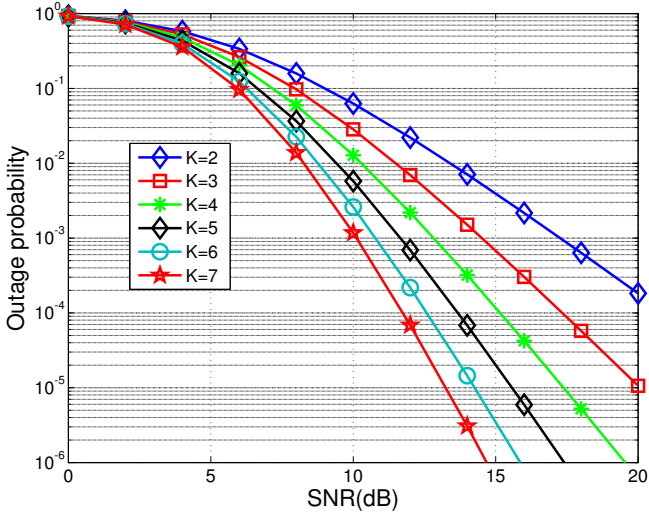}
\caption{Outage probability with $K$ relays }
\label{fig4}
\end{center}
\end{figure}
\begin{figure}[h]
\begin{center}
\includegraphics[width=3in]{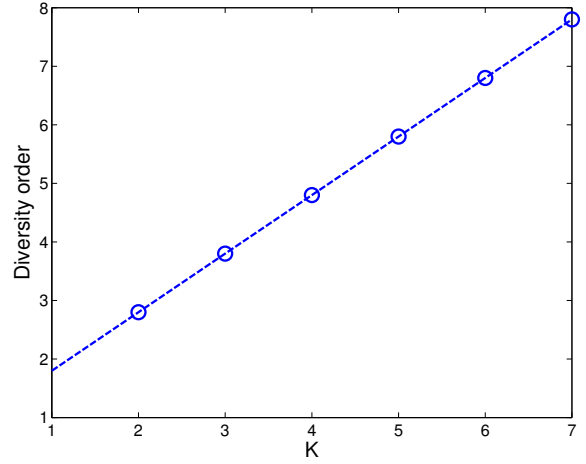}
\caption{Diversity order with different $K$}
\label{fig5}
\end{center}
\end{figure}
\begin{figure}[h]
\begin{center}
\includegraphics[width=3.2in]{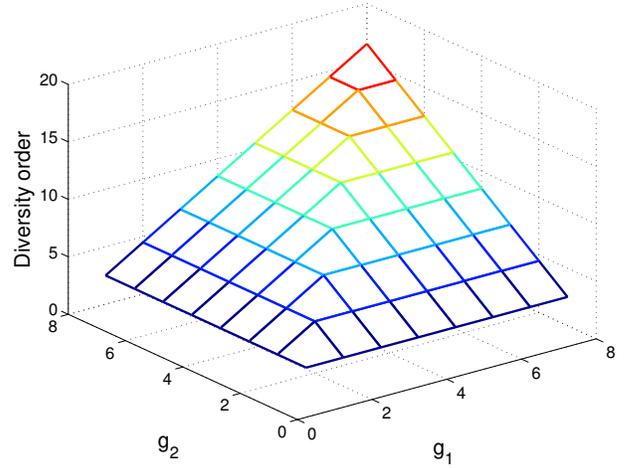}
\caption{Diversity order with different fading parameters}
\label{fig6}
\end{center}
\end{figure}

Fig. \ref{fig4} illustrates the theoretical results for the outage probability of opportunistic DF-AF scheme with different number of relays, $K$.
In the computation, we set $m_0=0.8$ and $m_{1i}=m_{2i}=1$.  When $m_{1i}=m_{2i}=1$, the diversity order is $d=m_0+K$ (See Fig. \ref{fig5}). From Fig. \ref{fig4}, we can clearly find that the number of relays impacts the slope of the curves. When there are more relays, the outage probability decreases more faster. In addition, the slopes of the curves in high SNR in Fig. \ref{fig4} are concordant with the diversity order illustrated in Fig. \ref{fig5}.
Fig. \ref{fig6} shows the diversity order with different channel fading parameters. We consider $2$ symmetric relays, i.e., $m_1=[g_1 ~g_1], m_2=[g_2 ~g_2]$. The fading parameter for the direct channel from the source to the destination is $m_0=0.5$. It can be observed that the diversity order is determined by the worse one in the source-relay channel ($g_1$) and relay-destination channel ($g_2$).
\par
\begin{figure}[h]
\begin{center}
\includegraphics[width=3.4in]{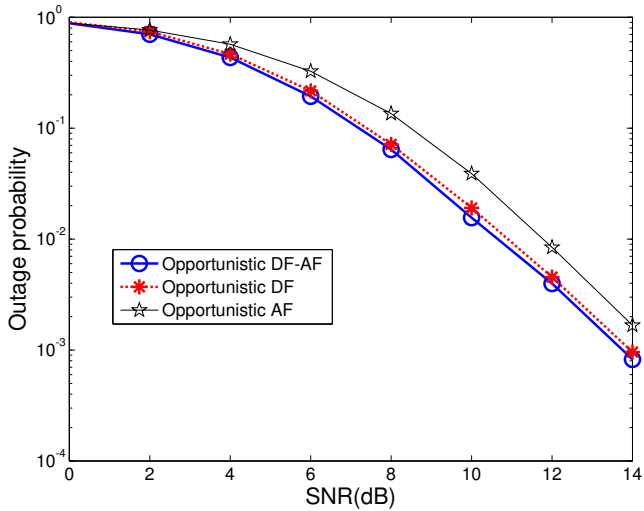}
\caption{Outage comparison in 3-relay networks when MRC is applied}
\label{fig7}
\end{center}
\end{figure}

To further demonstrate the advantages of the opportunistic DF-AF scheme, we show the outage performance of the three schemes when MRC is used in Fig. \ref{fig7}. The fading parameters are set as $m_0=0.5,m_{1}=[1 ~1 ~2],m_{2}=[1 ~1 ~1]$. Notice that the opportunistic DF-AF scheme has the best outage performance, which verifies the proposed theoretical analysis.
\begin{figure}[h]
\begin{center}
\includegraphics[width=3.4in]{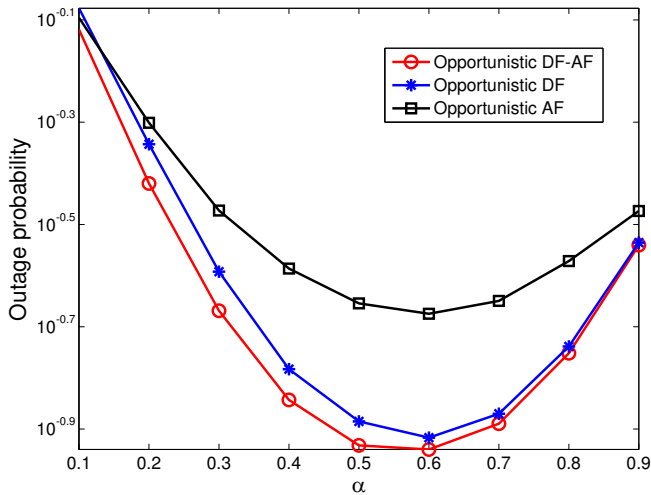}
\caption{Outage comparison with different power allocations between the source and the relay in 3-relay networks, the total power $P_s+P_r=10~dB$ and the fading parameters are chosen as $m_0=0.5,m_{1}=[1 ~1 ~2],m_{2}=[1 ~1 ~1]$}
\label{fig8}
\end{center}
\end{figure}
In Fig. \ref{fig8}, we consider different power allocations between the source and the relay. Define $\alpha=P_s/(P_s+P_r)$ as the power allocation coefficient, where $P_s$ and $P_r$ are the transmit power at the source and the relay, respectively. In the simulation, MRC is applied at the destination, and we set the AWGN variance $N_0=1$. From Fig. \ref{fig8}, we can see that the opportunistic DF-AF scheme outperforms the other two schemes. The advantages are obvious when $\alpha=0.2,...,0.7$, and the opportunistic DF-AF scheme has almost the same performance as the opportunistic DF scheme when $\alpha=0.8,0.9$. In addition, from the curve for the opportunistic DF-AF scheme, we can notice that with the increase of $\alpha$, the outage decreases at first and then increases. This can be explained as follows: When the source power increases, the relay has larger probability to correctly decode the source message, and then DF will be utilized with larger probability. Thus, the curves of the DF-AF scheme and DF scheme approach while we increase $\alpha$. The equivalent SNR of the relay path is given by (\ref{eqivelent}). When the source power is low (i.e., $\alpha$ is small), the relay could not decode the source messages with high probability, equivalent SNR of the relay path is approximated by $\frac{\gamma_{1i}\gamma_{2i}}{\gamma_{1i}+\gamma_{2i}+1}$. In this case, the increase $\alpha$ will result in the increases of $\gamma_{1i}$ and decrease of $\gamma_{2i}$. Observe that $\gamma_{1i}$ increase from a small number and $\gamma_{2i}$ decreases from a large number, i.e., $\gamma_{1i}$ and $\gamma_{2i}$ approach each other. Consequently, $\frac{\gamma_{1i}\gamma_{2i}}{\gamma_{1i}+\gamma_{2i}+1}$ increases. Thus, we have better outage performance. However, when the source power goes beyond a threshold, the relay has high probability to decode the source messages, and
the equivalent SNR of the relay path is determined by the SNR of the second hop $\gamma_{2i}$. The increase of $\alpha$ means the decrease of the relay power, i.e., the decrease of $\gamma_{2i}$. So we have higher outage probability. Finally, from the figure and the analysis, we can guess that an optimal $\alpha$ exists.
\section{Conclusion}
In this paper, we investigate the opportunistic DF-AF selection relaying scheme in wireless networks.
The optimal relay selection for maximizing received destination SNR is studied. We analyze the outage probability over Nakagami-$m$ fading channels,
and a closed-form solution is obtained. The coding gain and diversity order are derived whereby the asymptotic analysis in high SNR.
We find that the diversity order depends on not only the relay number but also the fading parameters. Moreover, we prove that the opportunistic
DF-AF selection relaying scheme outperforms both the opportunistic DF scheme and opportunistic AF scheme in terms of the outage performance. Finally,
the numerical results verify our analysis. In addition, simulations demonstrate that the power
allocation between the source and the relay plays an important role on the performance. We will investigate the the power
allocation to further improve the performance in the future works.
\ifCLASSOPTIONcaptionsoff
  \newpage
\fi

\end{document}